\documentclass[preprint]{seismica}
\usepackage{booktabs, multirow}
\title{Locating the Nordstream explosions without a velocity model using polarization analysis}
\shorttitle{Locating Nordstream explosions from seismic polarization} 

\author[1]{S. C. St\"ahler
	\orcid{0000-0002-0783-2489}
	\thanks{Corresponding author: mail@simonstaehler.com}
}
\author[1]{G. Zenh\"ausern
    \orcid{0000-0001-9401-4910}
}

\author[2]{J. Clinton
	\orcid{0000-0001-8626-2703}
}
\author[1]{D. Giardini
	\orcid{0000-0002-5573-7638}
}
\affil[1]{Institute for Geophysics, ETH Zürich, Switzerland}
\affil[2]{Swiss Seismological Service, ETH Zürich, Switzerland}

\credit{Conceptualization}{S. C. St\"ahler}
\credit{Software}{G. Zenh\"ausern}
\credit{Formal Analysis}{S. C. St\"ahler, G. Zenh\"ausern}
\credit{Resources}{J. Clinton}
\credit{Writing - original draft}{S. C. St\"ahler}
\credit{Writing - Review \& Editing}{S. C. St\"ahler, G. Zenh\"ausern, J. Clinton, D. Giardini}
\credit{Visualization}{G. Zenh\"ausern}
\credit{Supervision}{D. Giardini}

\begin{document}
	
\makeseistitle{
	\begin{summary}{Abstract}
	    The seismic events that preceded the leaks in the Nordstream pipelines in the Baltic Sea have been interpreted as explosions on the seabed, most likely man-made. We use a polarization-based location method initially developed for marsquakes to locate the source region without a subsurface velocity model. We show that the 2 largest seismic events can be unambiguously attributed to the methane plumes observed on the sea surface. The two largest events can be located with this method, using 4 and 5 stations located around the source, with location uncertainties of 30km and 10x60km.
	    We can further show that both events emitted seismic energy for at least ten minutes after the initial explosion, indicative of resonances in the water column or the depressurizing pipeline.
	\end{summary}
	 \begin{summary}{Zusammenfassung}
	    Die Lecks in den beiden Nordstream Pipelines wurden von zwei signfikanten Seebeben begleitet. Der Charakter dieser Seebeben spricht gegen einen tektonischen Prozess und für eine Explosion, gefolgt von schneller Dekompression des Gases. Wir verwenden eine Polarisationsanalyse, die die Richtung der Bodenbewegung analysiert, um die Beben zu lokalisieren. Diese Methode wurde ursprünglich entwickelt, um das Epizentrum von Beben auf dem Mars mit einem einzelnen Seismometer zu bestimmen. Wir zeigen, dass mithilfe von 5 Stationen um die westliche Ostsee die beiden Explosionen sicher den an der Oberfläche beobachteten Methan-Plumes zugeordnet werden können. Darüber hinaus können wir zeigen, dass auf die Explosionen ein mindestens zehnminütiger energie-reicher Dekompressionsprozess folgte. Mehrere Resonanzfrequenzen in diesem Signal deuten auf Reverberationen in der Wassersäule oder der geplatzten Leitung hin.
	 \end{summary}
	\begin{summary}{Non-technical summary}
		The leaks in the Nordstream pipelines, which transport natural gas from the Siberian gas fields to central and western Europe have been accompanied by seismic events consistent with underwater explosions. Seismic network operators located these explosions using the arrival times of different seismic wave types (P-waves, S-waves), that travel with different velocities. However, these velocities depend on the geological structure of a region and are often not well known, specifically in locations without many earthquakes. We therefore apply a method that uses the polarization, i.e. the direction, in which the ground is moving to determine the direction towards the seismic events. Using 5 stations around the Western Baltic Sea, we show that the two seismic events are located next to the observed gas leaks. We also show that the seismic events consisted of an initial explosion followed by an at least ten minute long process near the source, likely related to the rapid decompression of the pipeline and sound reflections between the sea floor and the surface.  
	\end{summary}
    }
	
	\section{Introduction}
	The catastrophic leakage events that occurred in the Nordstream 1 and 2 natural gas pipelines in the Baltic Sea on 26 October 2022 generated global interest due to their significance for the European gas supply and the relationship between the Russian Federation and the Western European nations at each end of the pipeline. Several hours after the first explosion shortly after 02:00 local time, a pressure drop was noticed at the German (western) end of the pipeline, Danish military intelligence reported large methane plumes at the sea surface and restricted the area to marine traffic. A second larger event occurred that evening shortly after 19:00 local time.  Despite the fact that the pipelines were not transporting any gas at the time of the leak, they were fully pressurized and thus several million tons of methane were released after the leak.
	A few hours after the initial leak, the Swedish national seismic network SNSN at Uppsala University \citep{lund_modern_2021} reported an earthquake of $M_{\mathrm{L}}=2.7$ near the now-confirmed location of the leak, based on picking arrival times of seismic waves \citep{snsn_swedish_1904}. The second event was also reported by SNSN as $M_{\mathrm{L}}=3.1$, close to the location of the second leak, clearly on the Nordstream 2 pipeline. Since the Baltic Sea is a region of very low seismicity \citep{grunthal_mw_2008}, it is plausible to identify these seismic events with the leaks and attribute them to an explosion, most likely deliberate.
	
	Seismic detection of man-made explosions is a task that dates back to the mid-20th century, when nuclear explosions were monitored by both super-powers. Coincidentally, the NORSAR array, which first reported the Nordstream quake, was set up precisely for this task. In the early period for seismology, event detection and location was not done using global networks but rather by single arrays that determined the back-azimuth and incident angle of seismic body waves by measuring the apparent horizontal slowness, i.e. the difference in arrival times, over a network of 10-100 km aperture. Such arrays included the NORSAR array, the LASA array in Montana, USA, the Warramunga array in Central Australia or the Gr\"afenberg array in South-Eastern Germany. The main motivation for using single arrays was that in the 1970s and 1980s, near real-time communication, as well as clock synchronization was not guaranteed in a global seismic network, so local arrays provided a more robust way to observe nuclear test signals from regional to teleseismic distances. Via observation of differential arrival time of seismic phases, the incident angle and the back-azimuth, an event could be located within the territory of a Nuclear power and attributed to a known test site, and its magnitude estimated to obtain the yield of a nuclear test. Improvement came with the installation of a global seismic network of digital recorders connected via satellite. The Comprehensive Test Ban Treaty Organization (CTBTO) was able to detect and localize nuclear test candidates by using arrival times at different stations and triangulate the source location. This however requires a reasonable model of seismic velocities. 
	In many regions of the world such models do not exist, coincidentally also in the Baltic Sea, a mostly aseismic region. The Baltic sea itself is a depression resulting from the widespread glaciation during the Weichsel ice age and undergoes a postglacial uplift of 4-5~mm/a in the Western part. The sea floor is covered with several 100~m of soft quaternary sediments but does show a surprising complexity, specifically south of Bornholm, where a system of graben faults points SW/NE, and the shallower Arnager block has exposed cretaceous bedrock at the surface. Hence the seismic velocity profile in the uppermost kilometers is complex \citep{ostrovsky_deep_1994-1, vejbaek_palaeozoic_1994}.
	
	We therefore apply a method for event localization that does not require a seismic velocity model and which was initially developed to locate seismic events on Mars. On Mars, we separately determine the direction of the marsquake as seen from a single seismic station (back-azimuth), based on the polarization of the main body waves: P and S. Since the P-wave is a compressional wave, its particle motion is in the direction of propagation, i.e. on a line pointing away from the epicenter. The S-wave is transversally polarized, i.e. orthogonal to the direction of polarization, which helps to determine the back-azimuth if the P-wave is not sufficient. The method is described in \citet{zenhausern_lowfrequency_2022}, where succesful application to teleseismic events on Earth is demonstrated. It is now routinely applied by the InSight Marsquake Service \citep[MQS, ][]{clinton_marsquake_2021, ceylan_marsquake_2022} to locate seismic events on Mars, where only a single seismometer operates and thus classical multi-station methods cannot be applied.

	\section{Method}
\begin{figure*}
	    \centering
	    \includegraphics[width=\textwidth]{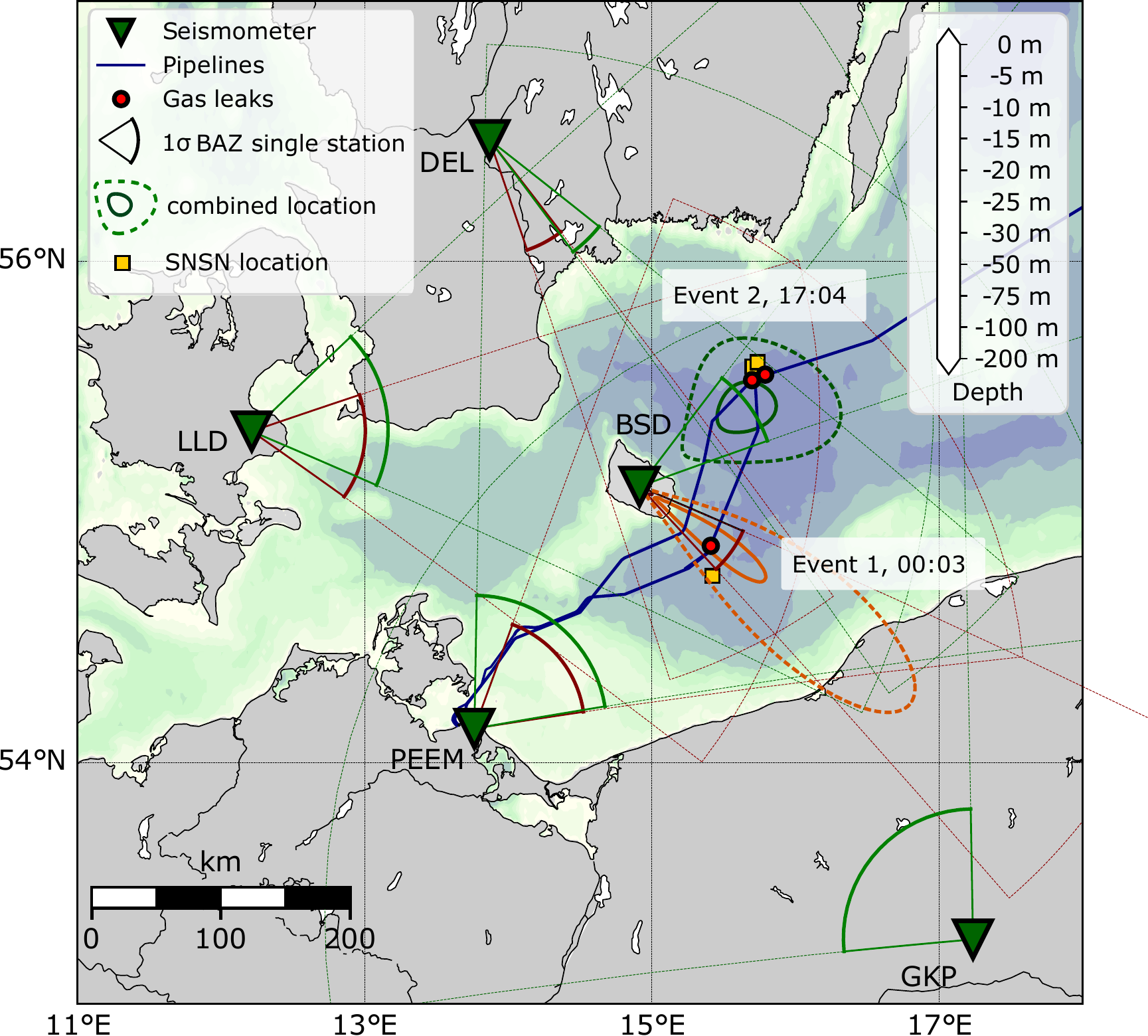}
	    \caption{Overview map of the Western Baltic Sea. We could obtain clear back-azimuths from 5 stations for event 1 (generally marked by reddish colors) and 4 for event 2 (marked by greenish colors). Event 1 is located less well, mainly due to the less favourable geometry of UP.DEL and DK.BSD, the two stations with best azimuth constraints. Additional stations like GE.RGN on the R\"ugen island (Germany) or DK.COP near Kopenhagen were tried, but had poorer azimuth constraints than neighbouring stations. We thus did not include them in this analysis and figure. The SNSN operates several more stations in Southern Sweden that might give additional constraints, but data from these was not publicly available at the time of writing.}
	    \label{fig:map}
	\end{figure*}

	We apply a complete polarization analysis of P and S body waves to determine the back-azimuth of seismic events. The three-component seismogram is transformed into time–frequency domain using a continuous wavelet transform \citep{kristekova_misfit_2006} to produce a time–frequency dependent complex spectral matrix. For each time–frequency pixel, the matrix is decomposed into eigenvectors to obtain information on the instantaneous polarization of the seismic signal. This method is based on the work of \citep{samson_pure_1983} and was first applied to seismic data by \citet{schimmel_use_2003, schimmel_polarized_2011}. 
	We use all open access stations from the European Integrated Data Archives \citep[EIDA][]{strollo_eida_2021} in a circle of 3~degrees (333 km) great-circle-arc distance around the reported position of the leaks. We download all HH? channels (high-sensitivity seismometer, typically sampled at 100~Hz) and correct the data to displacement. We then manually scan the data of days 2022-09-25 and 2022-09-26, i.e. the 2 days before the leaks were reported, for signals of nearby, high-frequency seismic events. As reported by SNSN, one event was found on 2022-09-26 17:03:50 UTC and a second event around 2022-09-26 00:03:24 UTC. Table \ref{tab:stations} has an overview over all stations on which the events were clearly detectable. For each of these stations, we identify a P-wave arrival window and apply our back-azimuth analysis to it in a 10~second time window starting 5~seconds before the arrival. See figure \ref{fig:pol_plot} for an example of our polarization analysis plot \citep{zenhausern_lowfrequency_2022}. Polarization plots for all investiated stations can be found in the supplement.
	To locate the event, we combine the azimuth dependent probability $p_i(\alpha)$ of multiple stations $i$ by multiplication
	\begin{equation}
	    p_{\mathrm{tot}}(\varphi, \theta) = \prod_{i=1}^N p_i(\alpha(\varphi, \theta)), 
	\end{equation}
	to obtain a probability density function for latitude $\theta$ and longitude $\varphi$. From this density function, a maximum likelihood value and an error ellipse is obtained and plotted in figure \ref{fig:map}.
	\begin{table*}[h]
	    \centering
	    \begin{tabular}{lllllll}
	    \toprule
	         & \multicolumn{2}{l}{Event 1} & Station & P arrival & S arrival & Back-azimuth [deg]\\
	        \cmidrule(lr){1-3}\cmidrule(lr){4-7}
	        Origin time & \multicolumn{2}{l}{00:03:24.5} & UP.DEL  & 00:03:55 & 00:04:25 & 153 [142-165] \\
            Latitude & \multicolumn{2}{l}{54.768} & PL.GKP  & 00:04:37 & 00:04:48 & -  \\
            Longitude & \multicolumn{2}{l}{15.431} & DK.BSD  & 00:03:32 & - & 125 [111-139] \\
            Magnitude & \multicolumn{2}{l}{2.7} & DK.LLD  & 00:04:00 & - & 100 [71-125] \\
            &  &  & KQ.PEEM  & 00:03:50 & 00:04:08 & 54 [20-81] \\
            \midrule
	         & Event 2A & Event 2B &  &  &  & \\
	        \cmidrule(lr){1-3}\cmidrule(lr){4-7}
	        Origin time & 17:03:50.4 & 17:03:58.5 & UP.DEL  & 17:04:15 & 17:04:37 & 135 [128-143] \\
            Latitude & 55.6 & 55.617 & PL.GKP  & 17:04:27 & - & 325 [265-2] \\
            Longitude & 15.71 & 15.745 & DK.BSD  & 17:04:03 & 17:04:11 & 55 [37-70] \\
            Magnitude & 3.1 & 3.1 & DK.LLD  & 17:04:30 & - & 85 [46-113] \\
            &  &  & KQ.PEEM & 17:04:20 & 17:04:45 & 33 [356-76] \\
            \bottomrule
	    \end{tabular}
	    \caption{Summary of key parameters from open accessible nearby stations. P and S-wave arrival times for each station with estimated back-azimuth. Back-azimuth uncertainty ranges are given in brackets. All times are on 2022-09-26 (UTC).}
	    \label{tab:stations}
	\end{table*}
	
	\section{Results}
\begin{figure*}
	    \centering
	    \includegraphics[width=\textwidth]{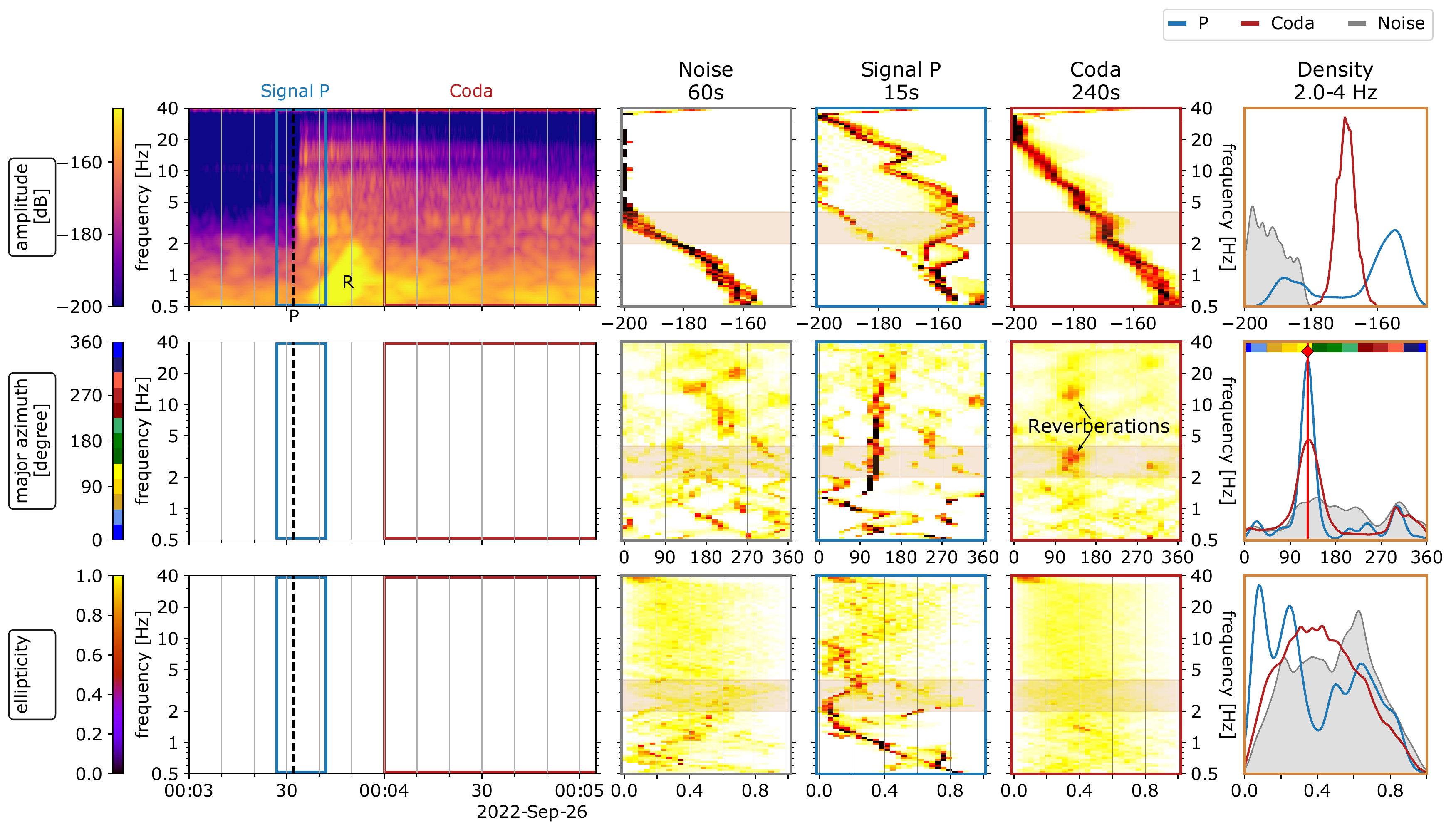}
	    \caption{Polarization analysis of event 1 (2022-09-22T00:02:30) for Station DK.BSD. The signals from this first event shows a clearly polarized P-wave up to 20 Hz, a Rayleigh wave between 0.5 and 2 Hz and sustained polarization after the event. The back-azimuth is estimated to $125\pm14$ degree from the P-wave. After the P-wave, polarization in the same direction is seen at 3.5 and 15~Hz, estimated as a continued pulsation at the source location.}
	    \label{fig:pol_plot}
	\end{figure*}

	We find clearly polarized P-waves at 4 (event 1) and 5 (event 2) stations in a distance range from 50 to 250~km. The clearest observation is on station DK.BSD located on Bornholm Island (see figure \ref{fig:pol_plot} for the first explosion), with a mostly marine path. For both events, the back-azimuth is constrained to less than 30\textdegree (see table \ref{tab:stations}. Together with the known geometry of the Nordstream pipelines and the locations of the methane plumes on the sea surface, an identification of the explosions would already be possible. The energy in the seismograms range from 0.2 to 40~Hz, with a clear P-wave but no obvious S-wave. Instead, a Rayleigh wave with clear elliptical polarisation arrives 10 seconds after the P. The overlap between S and Rayleigh is consistent with other quakes in distances of 50-100~km. The signal has an overall duration of at least 10 minutes before falling to pre-event noise levels. The Swedish National Seismic Network reported two separate explosions for the second seismic event, separated by 8 seconds, which we find to be consistent with the observation that the second pulse has the same polarisation attributes as the first.
	
	The second-closest station KQ.PEEM in Peenem\"unde, Germany, in 100 (event 1) and 150~km (event 2) distance has clearly visible signals as well. Both P and S-arrivals are visible, but back-azimuths are less constrained (60\textdegree uncertainty for event 1, 80\textdegree for event 2). The reduced amplitude is possibly due to the extended shallow sea over half of the distance to the events. The third station, UP.DEL in Southern Sweden, is significantly clearer in signal and shows a comparable back-azimuth constraint to DK.BSD. Surprisingly, this works even for the first event, which is located behind the Bornholm island as seen from the station. The fourth station, DK.LLD, shows a similarly bad constraint as KQ.PEEM, which is plausible given a low amplitude and paths that cross the Bornholm island and the lands of Southern Sweden.
	A signal is visible on several other openly accessible stations in Germany, Denmark, and Sweden, but the polarisation analysis did not obtain any additional constraints on the source locations.
	Multiplication of the probability density functions for all stations results in source regions close to the reported leaks. The first event has a very elongated uncertainty ellipse. For this event, the stations DK.BSD and UP.DEL are almost located in a line. The actual leak is located inside the $1\sigma$ region. For the second event, the stations are better positioned to constrain the location of the event very closely. The actual leak location is just outside the $1\sigma$ region, mainly due to the broad uncertainty from DK.BSD. 
	
	On stations DK.BSD, which is the closest station to either event, and UP.DEL, we find sustained polarization after the first event. This is a clear indication that the signal duration is not caused primarily by scattering but that seismic energy was radiated from the source over an extended period, at least 15 minutes. An explanation for this could be continued release of gas under high pressure. Two peaks are seen at 3.5 and 15~Hz. Assuming a speed of sound of 1450~m/s (typical for 15\textdegree C), this would correspond to wave lengths of about 95 and 400~meter. The water depth at the source is 70~meter, so this suggests that the 15~Hz signal could be an actual reverberation within the water column, while the 3.5 Hz signal is more likely an effect of the leak itself, potentially the Minnaert resonance of rising gas bubbles.

	\section{Conclusion}

	The analysis of P-wave polarization on the signal of the Nordstream pipeline explosions shows the strong potential of the method for a model-agnostic location of seismic events. We clearly identify both leaks with the separate seismic events. Location uncertainties from 4 and 5 stations' polarization were larger than those based on travel time methods, but the latter used significantly more stations. As opposed to travel time methods, our approach does not need a velocity model, is robust against timing errors on stations and can easily be started from a single station, as soon as data is available there.
	
	Both events show an absence of strong S-waves, consistent with a mostly isotropic source, such as an explosion. The closest station, DK.BSD on Bornholm, shows a clearly polarized coda, indicative of an ongoing source process over at least 10 minutes with several strong resonant peaks. This documents that polarization analysis of a small number of seismometer located onshore has the capability to locate and characterize seismic events in the water column.
	\begin{acknowledgements}
		Thank all relevant parties and acknowledge funding sources, if any.
		Seismic data were handled with ObsPy \citep{krischer_obspy_2015}. Calculations in Python were done with NumPy \citep{harris_array_2020} and SciPy \citep{virtanen_scipy_2020}, and the results were visualized with Matplotlib \citep{hunter_matplotlib_2007} and basemap.
		Seismic Data was collected using obspyDMT \citep{hosseini_obspydmt_2017} from EIDA (https://www.orfeus-eu.org/data/eida/). We recognise the following networks for providing data:
		Network DK, GE \citep{geofon_data_centre_geofon_1993}, PL, UP \citep{lund_modern_2021} and KQ \citep{christian_albrechts_-_universitat_zu_kiel_kiel_2017}
	\end{acknowledgements}
	
	\section*{Data and code availability}
	The polarization code is available on github 
	
    \section*{Competing interests}
    The authors have no competing interests.

\end{document}